# Predicted Trajectory Guidance Control Framework of Teleoperated Ground Vehicles Compensating for Delays

Qiang Zhang, Zhouli Xu, Yihang Wang, Lingfang Yang, Xiaolin Song, Zhi Huang

*Abstract*—Maneuverability and drivability of the teleoperated ground vehicle could be seriously degraded by large communication delays if the delays are not properly compensated. This paper proposes a predicted trajectory guidance control (PTGC) framework to compensate for such delays, thereby improving the performance of the teleoperation system. The novelty of this PTGC framework is that teleoperators' intended trajectory is predicted at the vehicle side with their delayed historical control commands and the LiDAR 3D point cloud of the environment, and then the vehicle is guided by the predicted trajectory. By removing the teleoperator from the direct control loop, the presented method is less sensitive to delays, and delays are compensated as long as the prediction horizon exceeds the delays. Human-in-the-loop simulation experiments are designed to evaluate the teleoperation performance with the proposed method under five delay levels. Based on the repeated measurement analysis of variance, it is concluded that the PTGC method can significantly improve the performance of the teleoperated ground vehicles under large delays(>200ms), such as the task completion time (*TCT*), deviation to centerline (*D2C*) and steering effort (*SE*). In addition, the results also show that teleoperators can adapt to smaller delays ($\leq 200$ ms), and the presented method is ineffective in such cases.

*Index Terms*—Delay compensation, Trajectory prediction, Guidance control, Teleoperation.

## I. INTRODUCTION

### A. Motivation

Although significant achievements have been made in recent years, fully autonomous driving remains challenging [1]. Humans still surpass machine intelligence in cognition. Benefiting from the merits of safety and cost, teleoperated vehicles are widely applied in dangerous areas and occasions unreachable by humans or too complex for an autonomous system, such as reconnaissance, route clearing, surveillance, and rescue [2]. In an unmanned ground vehicle (UGV) teleoperation system, the teleoperator watches the video feedback that is captured by onboard cameras and transmitted via wireless communication, and takes proper steering, braking and throttle operations like driving on a simulator.
A typical teleoperation framework is shown in Fig. 1, which is a Direct Control (DC) framework. The driver station and vehicle are spatially separated, and the remote vehicle is controlled directly by the teleoperator. Connections between vehicle and operator are realized by transmitting the teleoperator's commands $c(t)$ (including steering, throttle and brake commands) and the feedback $s(t)$ (including vehicle states, video feedback, etc.) through the wireless network. There exist control delays $t_{d1}$ and feedback delays $t_{d2}$ during transmission. This round-trip delay of $t_{d1} + t_{d2}$ leads to a temporal desynchronization between the operator's control actions and the observation of its corresponding vehicle response. When the delay is slight, human operators can adapt to delays by predicting the outcome of operations. However, when this delay is large, the human's adaptability to delay decreases due to high cognitive workloads resulting from a lack of clear correspondence between input and output [3], [4]. Our tests on a long-distance low-latency graph transmission system in an industrial district show that the typical latency ranges from 0.18s to 0.55s depending on the radio interference and terrain. When the input, i.e., video feedback, suffers from a considerable time delay (>200 ms) [5], the performance and stability of the teleoperation system could be degraded. To cope with the detrimental effects of large delay, a simple, effective method is to slow down the speed or adopt a move-and-wait strategy. Nevertheless, low efficiency is not acceptable in most cases.

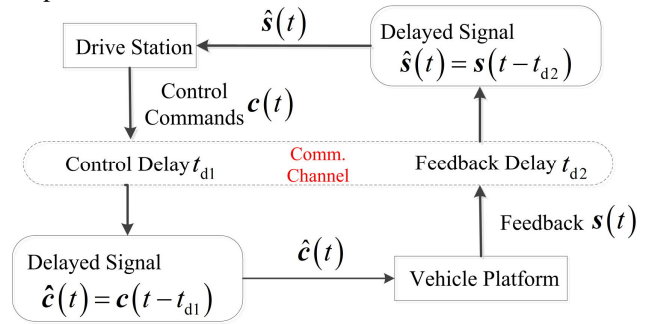

**Fig. 1.** Delays in a teleoperation system

This paper proposes a novel supervisory control framework for the ground vehicle teleoperation system to compensate for delays. The guidance trajectory is predicted based on the 3D point cloud in a bird's eye view (BEV) and driving commands. Then the teleoperated vehicle is directly controlled by a tracking controller to follow the predicted trajectory. The delay is compensated as long as the prediction horizon is greater than the delay.

### B. Related work

Since the latency increases the operator's cognitive workload [6], the principle of reducing cognitive workload is to maintain a correlation between commands issued by the operator and the expected result of those commands [7]. The predictive display aims to compensate for delays by predicting the vehicle and operator motion. In a predictive display solution, the vehicle response that is likely to result from the current operation of the operator is predicted and displayed immediately to help the operator receive feedback regarding their control actions. Brudnak [5] adopted a feed-forward vehicle model as a high-fidelity state estimator to predict the vehicle response. Graf [8] presented a curvature model using both the teleoperator's inputs and the vehicle states to calculate the trajectory curvature. The above vehicle model-based methods' performance depends on the models' accuracy. Dybvik [9] studied the effect of using a simple predictive display on performance and the operator's workload. Results from 57 participants showed a significant 20% improvement with the help of the predictive display. However, accurate acquisition of vehicle dynamics is challenging. Solutions that do not require the knowledge of vehicle dynamics have been considered in pursuit of robustness. Zheng [10], [11]proposed a model-free predictor to compensate for communication delays. In Zheng's study, the predictor is a first-order time delay system whose parameters are designed based on the stability analysis and the frequency domain performance analysis of coupling errors, so the presented method is sensitive to the delay and the frequency characteristic of coupling errors. Zheng's experiments show that human operators are affected more by the asynchrony between the generating steering commands and monitoring the subsequent vehicle heading than by the asynchrony between controlling and monitoring the vehicle's longitudinal speed. Zheng [12] further proposed a blended architecture for the vehicle heading prediction by combining the performance benefits of a model-based method with the robustness benefits of a model-free prediction scheme. To improve the situational awareness of the teleoperation system, Jung [13] developed head-mounted displays combined with a predictive display compensating for bidirectional network and operation delays to afford immersive 3D visual feedback.

Delay in the control loop results in deteriorated performance and instability. The robust control strategies against time delay have been intensively studied for teleoperation systems. Methods including Lyapunov function-based approaches [14], delay estimation techniques [15] and heuristic algorithms [16]– [18] are proven to be effective. Most studies focus on the bilateral teleoperation system to maintain stability and transparency. While for a ground vehicle teleoperation system, the main objective is to follow the operator's intention stably and avoid collisions. To deal with time-varying internet delay, Thomas [19] designed an adaptive Smith Predictor, which combined a delay estimation technique based on characteristic roots of delay differential equations to measure the delay with an adaptive Smith Predictor. In Thomas' approach, the teleoperator is modeled as a part of the control loop. Therefore, the variation of the teleoperator's response characteristics would affect the control system performance.

Adding autonomy capabilities to the teleoperated vehicle is an alternative approach handling delays. Studies verified that cooperative control could improve the performance and safety of unmanned ground vehicles [20], [21]. Cooperative control is classified into two categories: shared control and supervisory control. The key to shared control is the distribution of control right between the human operator and machine intelligence [22], [23]. Storms [22] presented an MPC-based shared control method. They found that communication delay's effect on safety has been improved considerably, while the control stability and the operator's workload are not discussed.

The supervisory control mitigates the sensitivity to delays by removing the operator from control loop. In a supervisory control system, the operator makes decisions based on environmental information and sets the global [24] or local guidance points/path to vehicles. The vehicle completes the maneuvering with its autonomous system. For some teleoperation applications, the environment is unknown or dynamic, so the global path guidance may not be applicable. Researchers paid more attention to the local path guidance model, also known as the point-to-go mode [25]. The operator needs to actively determine guidance points without decision support, which results in a significant cognitive workload. The vehicle speed fluctuates if the teleoperator cannot pick the guidance point timely. Zhu [26] proposed a method to generate candidate guidance points with the local perception information, decreasing the workload of picking the adequate guidance point. The main problem of the local guidance point/path-based approach is that teleoperators can hardly pick collision-free waypoints or paths due to the complexity of the driving environment and insufficient field feedback. Schitz [27] proposed an interactive corridor-based path planning framework. In Schitz's research, the human operator manually specified a corridor towards the destination in advance, and the vehicle planned a collision-free path in the specified corridor.

The above studies have addressed delay issues effectively. From the human perspective, experienced drivers get used to the normal driving manner, i.e., gazing at the area of interest, turning with a steering wheel, accelerating/decelerating with a pedal. Therefore, human operators using the normal driving manner could achieve better performance, such as stability and efficiency, than others due to lower cognitive workload. Predictive display mode seems preferable at this point. However, uncertain vehicle dynamics decrease the prediction accuracy and add to the workload of handling a vehicle. Although supervisory control is less insensitive to delays and vehicle dynamics since the autonomous system, instead of the teleoperator, is responsible for vehicle dynamics control, the pick-and-go mode could increase the cognitive burden. This study aims to develop a novel supervisory control framework to combine the merits of supervisory control and normal driving manner, in which the human teleoperates the vehicle as usual as normal driving, and the vehicle automatically follows the intended path compensating for delays.

*C. Contribution*

This paper proposes a predicted trajectory guidance control (PTGC) framework for teleoperated ground vehicles, aiming to improve the maneuverability and drivability of teleoperated vehicles under large delays. The proposed method uses a deep learning model to predict the operator's intended future trajectory and a tracking controller to follow this predicted trajectory. By removing the teleoperator from the control loop, the presented method is insensitive to delays while retaining the teleoperator's main authority over the vehicle by allowing the vehicle to drive as the driver intended.

The main contributions of this paper are summarized as follows:

1) A novel predicted trajectory guidance control (PTGC) framework is proposed to compensate for time delays, which reduces the teleoperator's cognitive workload and improves system performance by using a normal driving manner.

2) A deep learning-based multimodal prediction model using the operator's history operations and LiDAR 3D point cloud to predict the operator's intended trajectory is designed. The model achieves accurate trajectory prediction within one second. Thus, a delay of less than one second is compensated effectively.

*D. Paper Organization*

The remainder of the paper is organized as follows. Section II summarizes the system structure of the PTGC framework. The deep learning-based multimodal trajectory prediction model is presented in Section III. Section IV describes the details of the trajectory tracking controller. The design of human-in-the-loop experiments, the experimental results and discussions are given in Section V. Finally, Section VI makes the conclusions.

## II. PREDICTED TRAJECTORY GUIDANCE CONTROL FRAMEWORK

We propose a predicted trajectory guidance control (PTGC) framework, as shown in Fig. 2, to compensate for delays and improve the control stability. The PTGC framework comprises two modules: trajectory prediction and trajectory tracking.

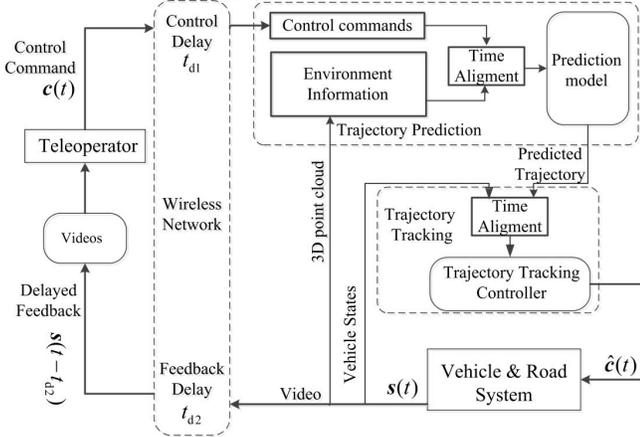

**Fig. 2.** Predicted trajectory guidance control framework

As discussed above, if the teleoperator controls the vehicle directly, the teleoperation system is sensitive to latency. In the PTGC framework, the teleoperator's commands are sent to the trajectory prediction module instead of directly to the vehicle and combined with environment information, i.e., 3D point cloud, to predict the teleoperator's intended trajectory. Since the teleoperator's control command $c(t)$ is the response to the vehicle-road system states $s(t - t_{d2})$. Therefore, the received control command $c(t - t_{d1})$ at the vehicle side is aligned with $s(t - t_d)$, here $t_d = t_{d1} + t_{d2}$, and fed into the intended trajectory prediction model. A deep learning-based multimodal trajectory prediction model generates the intended future trajectories with the prediction horizon greater than the total time delay. Compared to predictive display, where only driving commands are fed into a vehicle model to generate the future trajectory, the 3D LiDAR point cloud is incorporated here. The reasons are as follows: (1) Constant Turn Rate and Acceleration (CTRA) model that takes driving operations as input can only predict an accurate trajectory in a short period; (2) drivable area implied in the point cloud help generate a collision-free and feasible trajectory.

The predicted trajectory acts as the guidance trajectory and is fed into the tracking module, which controls the vehicle directly. The predicted trajectory is the outcome of input at time $t - t_d$, so the first $t_d$ of the predicted trajectory is truncated and then aligned with $s(t)$. A tracking controller outputs steering commands $\hat{c}(t)$ to vehicle according to the error between the actual and predicted trajectory. The trajectory tracking performance only depends on the tracking controller and is irrelevant to the time delay and the teleoperator's proficiency in driving skills.

Advantages of this PTGC framework are that, by predicting the intended trajectories, deploying the tracking controller at the vehicle side and removing the teleoperator from the control loop, the presented method is less sensitive to delays and vehicle dynamics, and the stability of trajectory following is improved.

## III. INTENDED TRAJECTORY PREDICTION USING LIDAR POINT CLOUD AND OPERATOR'S COMMANDS

In the proposed method, the vehicle is guided by the teleoperator's intended trajectory that is not picked directly by the operator but predicted with the operator's control command and environmental 3D point cloud. Our goal is to predict a future guidance trajectory over the next *T* time steps.

Intended future trajectory prediction can be considered a sequence-to-sequence problem. LSTM is with the ability to handle time sequence issues, so we construct trajectory prediction models based on LSTM networks. However, the general LSTM network can only predict one trajectory sequence and cannot perform multimodal prediction for the uncertainty of the operator's intention, which is prone to degradation of prediction accuracy. We proposed a method combining LSTM and multimodal prediction methods to address these problems, in which the Resnet is used to encode the context feature and LSTM is used to encode the motion feature.

*A. Trajectory prediction model*

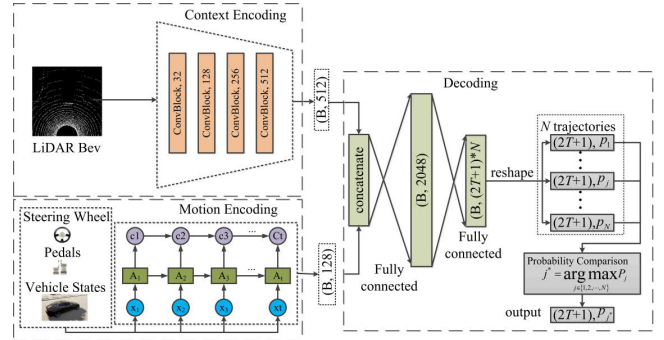

**Fig. 3.** Framework of the prediction model

The proposed prediction model is illustrated in Fig. 3. This model consists of three modules, i.e., motion encoding module, context encoding module and decoding module.

The motion encoding module is an LSTM network encoding control commands and vehicle states that implies the operator's intention. We predict the trajectory of the future $T$ time steps with information in the past $T_h$ time steps. The historical vehicle states are denoted as

$$\boldsymbol{S_t} = [\boldsymbol{s}_{t-T_h}, \cdots, \boldsymbol{s}_{t-i}, \cdots, \boldsymbol{s}_t] \quad (1)$$

where $\boldsymbol{s}_{t-i} = [x_{t-i}, y_{t-i}, v_{t-i}, \theta_{t-i}]$. $(x_{t-i}, y_{t-i})$, $v_{t-i}$, and $\theta_{t-i}$ are the position, velocity, and heading at the time step $t-i$, $i \in (1, \cdots, T_h)$, respectively.

The historical control commands are denoted as

$$\boldsymbol{C_t} = [\boldsymbol{c}_{t-T_h}, \cdots, \boldsymbol{c}_{t-i}, \cdots, \boldsymbol{c}_t] \quad (2)$$

where $\boldsymbol{c}_{t-i} = (\delta_{t-i}, Th_{t-i}, Br_{t-i})$. $\delta_{t-i}$, $Th_{t-i}$, and $Br_{t-i}$ are the steering, throttle, and brake commands at the time step $t-i$, respectively.

LSTM encoding module is to obtain the motion feature vector:

$$\boldsymbol{M_t} = LSTM(\boldsymbol{W} \cdot (\boldsymbol{S_t}, \boldsymbol{C_t}) + \boldsymbol{b}) \quad (3)$$

where the function $LSTM(\cdot)$ represents the input-output function of the LSTM, $\boldsymbol{W}$ and $\boldsymbol{b}$ are the weights and bias of LSTM, respectively.

The Context encoding module uses a Resnet network [28] to encode contextual constraints. The surrounding environment is described by the 3D point cloud. The point cloud in the range of $32\,\text{m} \times 32\,\text{m} \times 5\,\text{m}$ (length × width × height) is converted into a binary image on a BEV grid [29]. The grid resolution is 0.125 m. Therefore, the size of the binary image is $256 \times 256$ pixels. The binary image is then divided into two channels, one for ground points and the other for non-ground points, and produces the pseudo-image denoted as $\boldsymbol{B_t}$. The vector of context features $\boldsymbol{E_t}$ is obtained by encoding $\boldsymbol{B_t}$:

$$\boldsymbol{E_t} = Resnet(\boldsymbol{B_t}) \quad (4)$$

The dimensions of $\boldsymbol{M_t}$ and $\boldsymbol{E_t}$ are 128 and 512, respectively. We concatenate the context and motion features to generate a 640-dimensional feature that is fed into the decoding module to obtain the predicted output $\boldsymbol{P}_\text{out}$:

$$\boldsymbol{P}_\text{out} = Decoding(cat(\boldsymbol{M_t}, \boldsymbol{E_t})) \quad (5)$$

Note that we are not to generate one trajectory but $N$ candidate trajectories and their corresponding probabilities, so the dimension of $\boldsymbol{P}_\text{out}$ is $(2T+1) \cdot N$.

$$\boldsymbol{P}_\text{out} = [\boldsymbol{P}_1, \boldsymbol{P}_2, \cdots, \boldsymbol{P}_j, \cdots \boldsymbol{P}_N] \quad (6)$$

where $\boldsymbol{P}_j = [(x_{t+1}^j, y_{t+1}^j), \cdots, (x_{t+k}^j, y_{t+k}^j), \cdots, (x_{t+T}^j, y_{t+T}^j), p_j]$. $(x_{t+k}^j, y_{t+k}^j)$ is the predicted waypoint of the $j$-th trajectory at the time step $t+k$, $p_j$ is the probability of $j$-th trajectory, $k \in (1, \cdots, T)$, $j \in (1, \cdots, N)$ and $\Sigma_{j=1}^N p_j = 1$. The candidate with the highest probability is chosen as the predicted trajectory.

*B. Loss function*

The loss function [30] is constructed for the multimodal prediction. Specifically, we use the binary cross-entropy loss of classification and $smooth\ L_2$ loss for the trajectory regression tasks to calculate the total loss $\mathcal{L}_{all}$.

The trajectory regression loss $\mathcal{L}_{re}$ is defined based on $Mixture\text{-}of\text{-}Experts$ loss:

$$\mathcal{L}_\text{re} = \Sigma_{j=1}^N I_{j=j^*} \Sigma_{i=1}^T \|w_i^j - w_i^{gt}\|_2 \quad (7)$$

where $w_i^j$ is the predicted waypoint of the $j$-th candidate trajectory at time step $t+i$, and $w_i^{gt}$ is the ground-truth position. $I_{j=j^*}$ is a selection function setting to 1 if $j = j^*$ is true and 0 otherwise. $j^*$ is the number of the closest trajectory to the ground-truth trajectory according to the trajectory distance function.

$$j^* = \underset{j \in \{1, \cdots, N\}}{\arg\min} \Sigma_{i=1}^T \|w_i^j - w_i^{gt}\|_2 \quad (8)$$

$\mathcal{L}_{class}$ is the classification cross-entropy loss defined as

$$\mathcal{L}_{class} = -\Sigma_{j=1}^N I_{j=j^*} \log p_j \quad (9)$$

The total loss $\mathcal{L}_{all}$ is the sum of the trajectory regression loss and classification loss.

$$\mathcal{L}_{all} = \mathcal{L}_{re} + \alpha \mathcal{L}_{class} \quad (10)$$

where $\alpha$ is the classification loss weight to balance classification and regression performance.

*C. Ablation Experiments and Error Analysis*

Ablation experiments were conducted to analyze the significance of input components in the proposed model. The driving simulation was conducted on a CARLA-based simulator to collect the dataset for model training. We recruited six volunteers to drive vehicles on the simulator for data collection. The sample rate is 20Hz, and the time step *st* is 0.1 seconds. The trajectory in the past 20 steps (2s) was observed, and motion in the next 5, 10, and 20 steps (0.5 s, 1 s, and 2 s) was predicted. The collected data were split by a sliding window, and 63897 records were acquired. The ratio of records for training, validation and testing was 3:1:1.

We use the CTRA model [31] as the baseline. Variants of the proposed model, i.e., M-model using motion features, C-model using context features and MC-model using both motion and context features, are devised for comparison. The performance was evaluated using the average deviation error (ADE) and the final deviation error (FDE). Results of ablation experiments are shown in Fig. 4 and Table I.

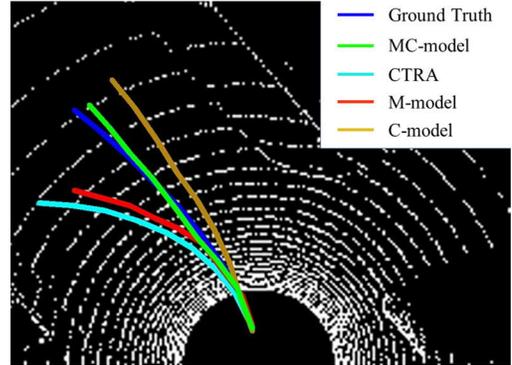

**Fig. 4.** Predicted trajectory using different model

TABLE I
RESULTS OF ABLATION EXPERIMENTS

| Time | 0.5 s | | 1 s | | 2 s | |
|---|---|---|---|---|---|---|
| Metric (m) | ADE | FDE | ADE | FDE | ADE | FDE |
| CTRA | 0.17 | 0.33 | 0.66 | 1.04 | 1.09 | 1.49 |
| M-model | 0.13 | 0.21 | 0.43 | 0.76 | 0.66 | 1.27 |
| C- model | 0.15 | 0.20 | 0.39 | 0.72 | 0.71 | 1.22 |
| MC-model | 0.13 | 0.18 | 0.24 | 0.54 | 0.56 | 0.86 |

We can find that the history command sequence and vehicle states are more significant for trajectory prediction than the current ones since M-model is better than the CTRA model. The motion feature and context feature are almost of equal importance. Combining motion and context features further benefits the trajectory prediction and is with the minimum ADE and FDE.

The distribution of large FDE errors for the prediction horizon of one second was further analyzed and presented in Table II. It shows that MC-model is with a smaller percentage of large error than the M-model and C-model. Especially, the error distribution of the MC-model is less than 1% in the range of error greater than 2m, which is much smaller than the other two models. It further proves that combining motion and context features is beneficial for trajectory prediction with a smaller error range. Therefore, the ablation experiments verified the feasibility of the proposed method. If not otherwise stated, the MC-model was adopted as the prediction model of the PTGC method.

TABLE II
DISTRIBUTION OF LARGE FDE ERRORS FOR THE 1S PREDICTION HORIZON

| Error | M-model | C-model | MC-model |
|---|---|---|---|
| error > 1 m | 9.4% | 8.5% | 5.8% |
| error > 1.5 m | 7.2% | 6.8% | 3.3% |
| error > 2 m | 4.2% | 3.1% | 0.6% |
| error > 2.5 m | 2.3% | 1.6% | 0.3% |
| error > 3 m | 0.9% | 0.7% | 0.1% |

## IV. TRAJECTORY FOLLOWING CONTROL

As mentioned in section II, the predicted trajectory is the outcome of input at time $t - t_d$, so the first $t_d$ of the predicted trajectory is truncated. As shown in Fig. 5, the predicted trajectory $\widehat{Tra}$ consists of two segments, i.e., the history segment to be truncated (dashed line) and the future segment (solid line) to be followed. The split point $w_{t_d}$ is the predicted waypoint at the predicted time step $t + t_d$. The Stanley control method [32] based on distance and heading angle errors at the point $w_{t_d}$ is adopted for trajectory tracking control. A simplified bicycle model with infinite tire stiffness is adopted in the design of the controller.

As shown in Fig. 5, the point $w_{td}$ is set as the preview point, $\theta_v$ is the vehicle heading angle, $\theta_p$ is the tangential angle of $\widehat{Tra}$ at $w_{td}$, and $e$ is the lateral deviation. The heading deviation is defined as

$$\theta_e = \theta_p - \theta_v \quad (11)$$

The steering control variable $\delta(t)$ can be obtained intuitively from the relative deviation of the vehicle position to the predicted trajectory, which contains the lateral deviation $e$ and the heading deviation $\theta_e$.

$$\delta(t) = \delta_e(t) + \delta_{\theta_e}(t) \quad (12)$$

If ignoring the lateral error, the direction of the front wheel is aligned with the tangential direction of its corresponding preview point and the heading component $\delta_{\theta_e}(t)$ is defined as

$$\delta_{\theta_e}(t) = \theta_e(t) \quad (13)$$

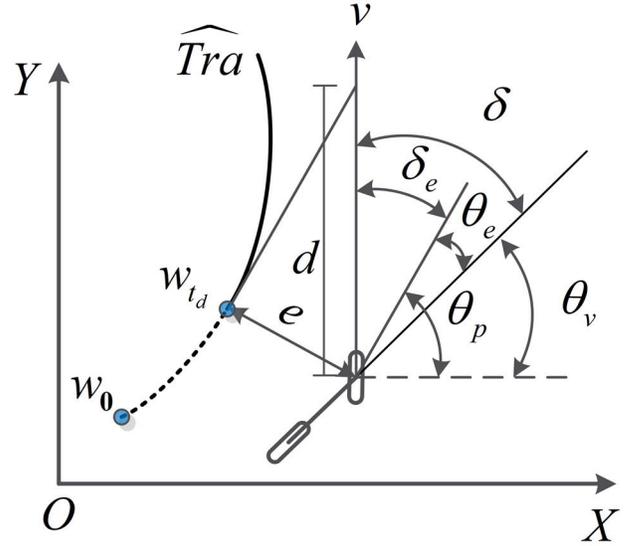

**Fig. 5.** Trajectory tracking control algorithm

Similarly, if not consider the heading deviation, the larger the lateral error is, the larger the front wheel steering angle is. Assuming that the expected vehicle trajectory intersects the tangent line of the preview point at a distance $d(t)$ from the front wheel, the lateral component $\delta_e(t)$ can be derived approximately from the geometric relationship if $\delta_e(t)$ is not large.

$$\delta_e(t) \approx \arcsin\frac{e(t)}{d(t)} = \arcsin\frac{ke(t)}{v(t)} \quad (14)$$

where we define $d(t)$ relating to the vehicle speed $v(t)$, i.e., $d(t)=v(t)/k$, and $k$ is a gain parameter greater than zero. The function $\arcsin(\cdot)$ produces a front-wheel deflection angle pointing directly to the trajectory to be tracked and is limited by the vehicle speed $v(t)$.

Considering the above two control components together in the steering angle control law $\delta(t)$, we have

$$\delta(t) = \theta_e(t) + \arcsin\frac{ke(t)}{v(t)} \quad (15)$$

Using a linear bicycle kinematic model, the change rate of the lateral error $\dot{e}(t)$ is given by

$$\dot{e}(t) = -v(t)\sin\delta_e(t) \quad (16)$$

where $\sin \delta_e(t)$ is known from the geometric relationship.

$$\sin \delta_e(t) = \frac{e(t)}{d(t)} = \frac{ke(t)}{v(t)} \quad (17)$$

$\dot{e}(t)$ is further expressed as

$$\dot{e}(t) = -ke(t) \quad (18)$$

We can get

$$e(t) = e(0)e^{-kt} \quad (19)$$

Thus, the lateral error converges exponentially to zero, and the parameter $k$ determines the convergence rate.

## V. EXPERIMENTS AND ANALYSIS

### A. Human-in-the-loop simulation platform

As shown in Fig. 6, a real-time driver-in-the-loop simulation platform was developed for teleoperation experiments and algorithm evaluation. The simulation platform consists of four parts, i.e., the drive station, the vehicle-road system, the controller and the communication network.

At the drive station, the operator uses a Logitech® G27 joystick for steering, braking and acceleration control, and a monitor for visual feedback. As shown in Fig. 7, the front view, vehicle speed and the predicted trajectory are shown on the monitor. Operators drive the vehicle based on this feedback.

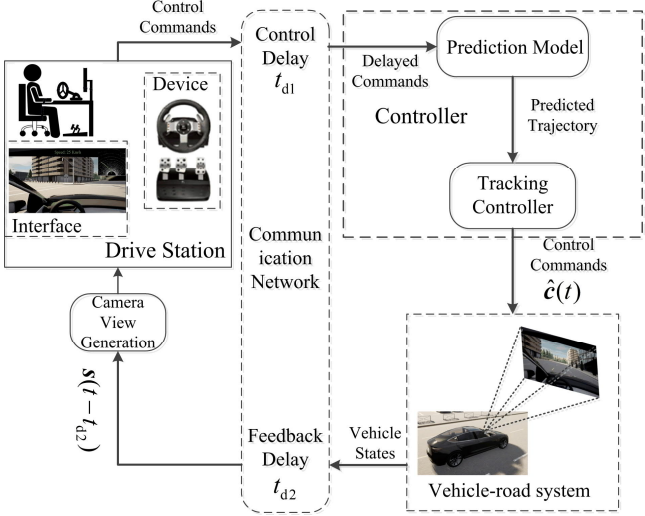

**Fig. 6.** Human-in-the-loop simulation platform for teleoperated ground vehicle system

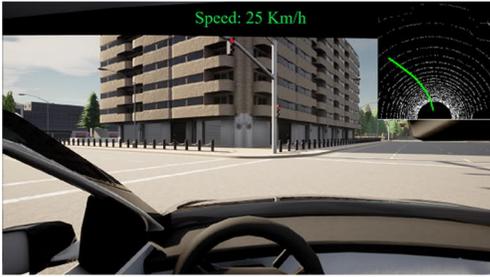

**Fig. 7.** Visual feedback displayed on the monitor

The vehicle-road system is simulated with a CARLA-based simulator. The CARLA simulator runs vehicle dynamics and physical world simulation, and outputs vehicle states and environment information, including the driver's view of the environment and the 3D LiDAR point cloud. The resolution of RGB camera is 800×600 pixels at a frame rate of 20 Hz. The LiDAR has 32 lines, scans 128,000 points per second, and outputs a laser point cloud at 20Hz.

The communication network is simulated by a ROS node. The control command and visual feedback are sent to the ROS node and queued in a first-in-first-out (FIFO) pipeline. The communication delay is realized by setting the depth of the FIFO pipeline greater than zero. Compared to the real teleoperation system, only the vehicle-road system dynamics and communication system are simulated, while the human-machine interface is almost identical. Therefore, the simulation platform can ensure the fidelity of the operator's response to delays.

*B. Test Road*

The test road is designed to evaluate the performance of the presented method under various delays. Due to the restriction of CARLA, the structural road scene was applied. As shown in Fig. 8, the 622m long closed-loop road features two-way four lanes and six turns with the radii ranging from 17 m to 45 m.

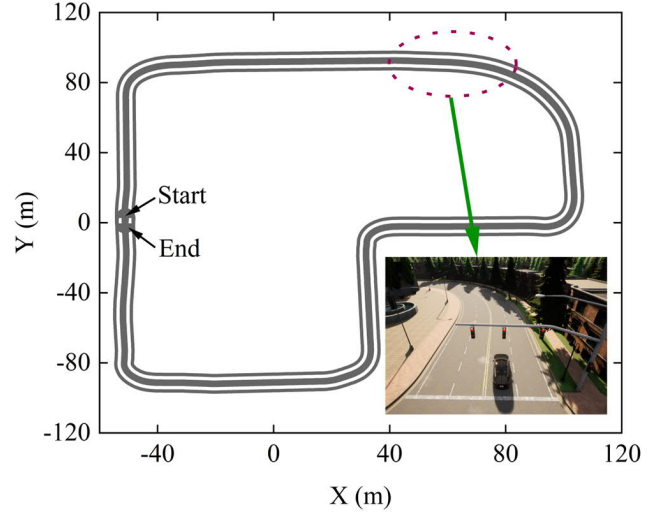

**Fig. 8.** Overview of test road

*C. Teleoperation tasks and performance metrics*

The study aims to improve the maneuverability and drivability of the teleoperated vehicle under large delays. We expect the proposed method to enable the vehicle to complete driving tasks as fast and safely as possible under different delay conditions. Therefore, the participants are required to operate the vehicle to complete the driving task as fast as possible while staying as close to the road centerline as possible to minimize tracking errors.

Three independent parameters, i.e., task completion time (*TCT*), deviation to centerline (*D2C*) and steering effort (*SE*), are used as performance metrics. *TCT* is defined as the time it takes for a participant to complete one loop of the driving task as "fast" and "smooth" as possible. The *D2C* is defined as the area between the actual vehicle track and the road centerline, indicating the magnitude of deviation to the centerline. These two metrics reflect the teleoperation system's longitudinal and lateral maneuverability performance, respectively. The lower value of metrics indicates higher performance. The *SE* is denoted by the average absolute steering angle, which characterizes the controllability of the teleoperated vehicle. Due to the detrimental effects of delay on teleoperated driving tasks, driving operations without timely visual feedback could result in oversteering and repetitive correction in the form of overdriving behavior. Less steering effort means more manageable and more comfortable control of the vehicle.

Teleoperated driving tasks under five delay levels ranging from 200ms to 1000ms with the interval of 200ms and applying two control frameworks, i.e., direct control (DC) and predicted trajectory guidance control (PTGC), are studied. The experiments follow a 5×2 within-subject factorial design and aim to determine how the delay magnitude and control method affect mobility and drivability. Including the zero-delay case as the baseline, eleven driving tasks are tested, and each task is repeated three times. Therefore, each participant needs to complete 33 runs. The task sequence is randomly scheduled to reduce the learning impact on individual scenes or one trajectory.

*D. Experimental procedure*

Nine people with an average age of 22 ± 3 years were recruited to participate in the experiments. They had a driver's license and at least one year's driving experience. All participants had a normal or corrected-to-normal vision and some experience driving in a virtual environment with a steering wheel and pedals (e.g., playing a virtual racing game) but no teleoperation driving experience in a delayed condition.

The whole testing process was divided into two sessions: the training session and the testing session. The training session served to help participants adapt to teleoperation under large latency based on the simulation platform. Participants were verbally informed of the test details, including the driving task and performance goals, i.e., completion time, deviation error and steering effort. Participants were asked to complete the driving task as quickly as possible but were not told which metric had a higher priority. Instead, it is up to them to adapt and adjust to the driving task. Once the training session had been completed, the test session began. Participants were asked to run 11 tasks in a randomized order during the testing session, and each task was repeated three times. A run was valid if the following events did not occur.
1) The vehicle ran off the road for 5 s.
2) Vehicle rollover.
3) The average speed is less than 18 km/h.

*E. Analysis methods*

A two-way RM-ANOVA was used to study the effects of two independent variables, i.e., delay level and control method, on *TCT*, *D2C* and *SE*. Here, two-way refers to two factors: delay level and control method.

The two null hypotheses for each metric were tested using an *F*-test based on the type III sum of squares and 95% confidence level. These null hypotheses are as follows:

(1) There is no significant difference in performance metrics when different control methods are used for teleoperation.

(2) No significant differences in performance measures exist between the different delay levels.

If the *F*-test indicated that at least one mean is different from the others (i.e., P < 0.05), Fisher's least significant difference method was used to identify the groups with pairwise significant differences in the means.

*F. Experimental results and discussion*

A total of 297 records were obtained, and one participant's result was discarded as the average speed was lower than 18km/h. Therefore, 264 valid records were used for experimental analysis.

A case (800 ms delay) study is shown in Fig. 9. Compared with the DC, the actual path is closer to the destined path when using the PTGC, and the steering intention of the driver can be recognized in advance before entering the intersection, thus controlling the vehicle steering as early as possible. Especially in the adjusting phase after the turning, the DC case has a longer overshoot, while using the PTGC reaches stability more quickly after turn, which means that using the PTGC has better maneuverability and stability.

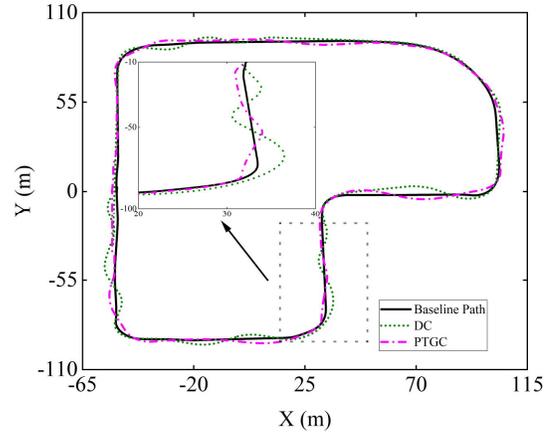

**Fig. 9.** Actual path under 800 ms delay

TABLE III
RM-ANOVA RESULT FOR THE METRIC *D2C*

| Factor | DF | F | P |
|---|---|---|---|
| Control | 1 | 30.63 | 0.000 |
| Delay | 4 | 28.61 | 0.000 |
| Control * Delay | 4 | 4.39 | 0.002 |

TABLE IV
RM-ANOVA RESULT FOR THE METRIC *TCT*

| Factor | DF | F | P |
|---|---|---|---|
| Control | 1 | 30.63 | 0.000 |
| Delay | 4 | 28.61 | 0.000 |
| Control * Delay | 4 | 4.39 | 0.013 |

TABLE V
RM-ANOVA RESULT FOR THE METRIC *SE*

| Factor | DF | F | P |
|---|---|---|---|
| Control | 1 | 30.63 | 0.001 |
| Delay | 4 | 28.61 | 0.000 |
| Control * Delay | 4 | 4.39 | 0.005 |

Two-way RM-ANOVA on each performance metric was conducted individually as a general linear model. The details of RM-ANOVA are shown in Tables III-V. With a significance level of 0.05, the *P* values of the *F*-test with respect to the factor of delay level are close to 0 and much smaller than 0.05 in all three RM-ANOVA tables, which indicates that the hypothesis of no significant difference between the different delay levels tested can be rejected with 95% confidence level. In terms of the effect of the control method, the P values for the metrics of TCT, D2C, and SE are 0.000, 0.000 and 0.001, respectively. All P values are smaller than 0.05. Thus, there is a significant difference in performance metrics when different control methods are used with a 95% confidence level. However, the *P* values of the factor Control*Delay for three metrics are 0.002, 0.013, and 0.004, respectively, indicating a significant interaction effect between the Control Method and the Delay Level. Therefore, we conducted a pairwise ANOVA comparison to determine whether the control method significantly affects the teleoperation performance at different delay levels. Results are shown in Fig. 10, where the '*' denotes the pairs with a statistically significant difference. For delays greater than 200 ms, the performance improvements in *D2C* and *TCT* metrics are significantly different using the PTGC relative to the DC, as shown in Fig. 10(a) and 10(b). The performance improvements in the *SE* metric are significant only when delays are greater than 400 ms, as shown in Fig. 10(c). All three metrics slightly worsen when the delay is not greater than 200 ms, indicating that humans can adapt to slight delay without assistance. The performance even decreases in the existence of assistance due to the prediction errors.

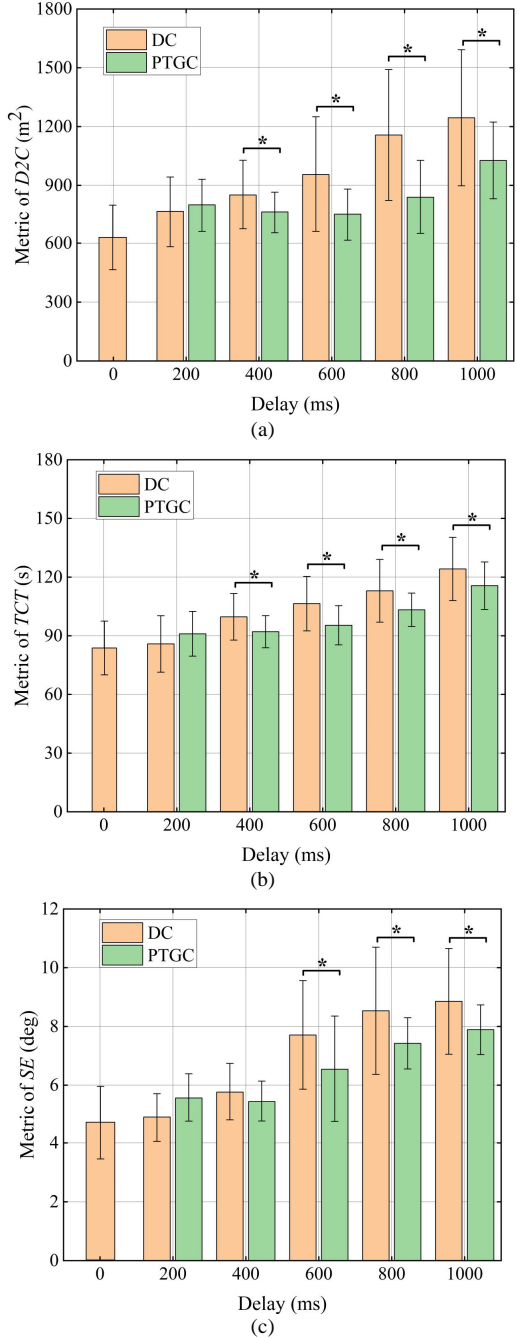

**Fig. 10.** Pairwise ANOVA comparison on *D2C*, *TCT* and *SE* at different delay levels

To find out why there exist performance differences under different delay levels using the PTGC, a one-way ANOVA was applied in DC cases to explore the performance differences to the baseline. The results are shown in Table VI.

TABLE VI
ONE-WAY ANOVA RESULTS FOR DIFFERENCES TO THE BASELINE USING DC METHOD

| Metric | Delay (ms) | Mean Difference | P |
|---|---|---|---|
| D2C | 0 | 200 | -133.464 | 0.433 |
| | | 400 | -222.104* | 0.008 |
| | | 600 | -326.271* | 0.000 |
| | | 800 | -528.021* | 0.000 |
| | | 1000 | -616.021* | 0.000 |
| TCT | 0 | 200 | -1.929 | 1.000 |
| | | 400 | -15.785* | 0.000 |
| | | 600 | -22.554* | 0.000 |
| | | 800 | -29.098* | 0.000 |
| | | 1000 | -40.285* | 0.000 |
| SE | 0 | 200 | -0.195 | 1.000 |
| | | 400 | -1.078 | 0.095 |
| | | 600 | -3.023* | 0.000 |
| | | 800 | -3.847* | 0.000 |
| | | 1000 | -4.168* | 0.000 |
| * indicates a significant difference | | | | |

All three performance metrics under 200ms delay are not significantly different from the zero-delay cases ($P < 0.05$), which indicates that human drivers can adapt to the low delays ($\leq 200$ ms) without compromising driving performance. Under 400 ms delay, the *SE* of DC cases is also not significantly different from the baseline ($P<0.05$). The reason could be that human drivers try to control the vehicle with minimal effort at relative low delay levels. The significance results at different delay levels in Table VI are consistent with those shown in Fig. 10, which indicates that the PTGC framework can result in remarkable performance improvements only when the DC framework is significantly affected by the delay. The reason is that the PTGC framework is based on the driver's intention prediction, and due to errors in trajectory prediction, its performance is always inferior to the baseline. To analyze the improvement quantitatively, results are further normalized using the average performance metrics of the zero-delay case as the benchmark [33]. The *D2C*, *TCT* and *SE* performance improvement of PTGC cases relative to DC cases at different delay levels is denoted as $P_{D2C}$, $P_{TCT}$, and $P_{SE}$, respectively. Referring to [34] and taking $P_{D2C}$ for example, the improvement is calculated by

$$P_{D2C} = \frac{|r_c - r_d|}{|r_d - r_0|} \qquad (20)$$

where $r_c, r_d$ and $r_0$ are the means of *D2C* of PTGC, DC and zero-delay cases, respectively.

Assuming each metric contributes equally to the overall performance $P_{\text{ove}}$, we get:

$$P_{\text{ove}} = \frac{P_{D2D}}{3} + \frac{P_{TCT}}{3} + \frac{P_{SE}}{3} \qquad (21)$$

Note that the overall performance is based on the significance analysis. If the performance improvement at a certain delay level is not statistically significant, the value is set to zero.

TABLE VII
PERFORMANCE IMPROVEMENT VS. DELAY LEVELS

| Delay | 400 ms | 600 ms | 800 ms | 1000 ms |
|---|---|---|---|---|
| $P_{D2C}$ | 41% | 59% | 60% | 35% |
| $P_{TCT}$ | 48% | 49% | 33% | 21% |
| $P_{SE}$ | 0 | 38% | 28% | 23% |
| $P_{ove}$ | 30% | 49% | 41% | 27% |

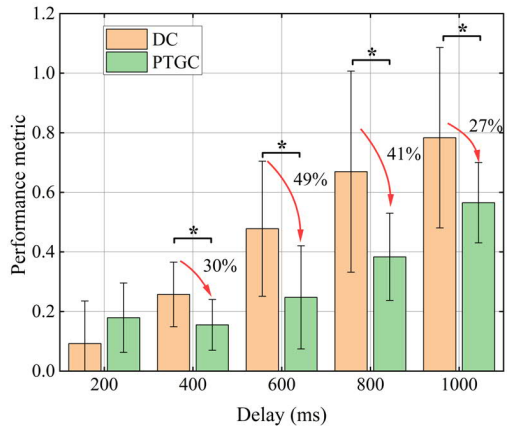

**Fig. 11.** Pairwise ANOVA comparison on the normalized overall performance metrics

The normalized improvement in the three metrics and the pairwise ANOVA comparison of the normalized overall performance metrics at different delay levels are shown in Table VII and Fig.11, respectively. It shows that under large delay levels, e.g., 400ms, 600ms, 800ms and 1000ms, the overall performance improvement of PTGC cases over the DC cases is 30%, 49%, 41% and 27%, respectively, and the overall performance improvement with the PTGC framework is statistically significant at these delay levels. The performance improvement decreases as the time delay increases when the delay is greater than 600ms. The reason is that the prediction error increases as the prediction horizon increases, and the prediction error is insignificant when the prediction horizon is less than 600ms. So, the overall improvement reaches its maximum at the delay level of 600ms.

VI. CONCLUSION

This paper proposes a predicted trajectory guidance control framework for teleoperation of ground vehicles, aiming to improve the maneuverability and drivability of teleoperated vehicles under delays. The control method is novel in that it uses a deep learning model to predict the teleoperator's driving intentions and intended trajectories at the vehicle side, and the vehicle is guided by the predicted trajectory using a closed-loop tracking controller. The advantage of this approach is that it removes the teleoperator from the closed-loop control system and reduces the sensitivity of the human driver to time delays. The performance of the proposed method is verified with a human-in-loop driving simulation at delay levels ranging from 200ms to 1000ms. Three performance metrics, i.e., *D2C*, *TCT* and *SE,* are used to evaluate the performance improvement. The results show that the proposed method improves maneuverability and drivability under delays>200ms. Under 600ms delay, the overall improvement is about 49%. However, there is no improvement for cases of delay ⩽ 200ms due to the human teleoperator's adaptability to small delays.